\documentclass{aa} % option [referee] to be inserted
\usepackage{graphicx}
\usepackage{natbib}
\usepackage[english]{babel}
\usepackage[T1]{fontenc}
\usepackage[latin1]{inputenc}
\usepackage[dvips]{epsfig}
\usepackage{amsmath}

\begin{document}

\title{Massive molecular outflows}

\author{H. Beuther\inst{1} \and P. Schilke\inst{1} \and T.K. Sridharan\inst{3} \and K.M. Menten\inst{1} \and C.M. Walmsley\inst{3} \and F. Wyrowski\inst{1,4}}

\institute{Max-Planck-Institut f\"ur Radioastronomie, Auf dem H\"ugel 69, 53121 Bonn, Germany \and Harvard-Smithsonian Center for Astrophysics, 60 Garden Street, MS 78, Cambridge, MA 02138, USA \and Osservatorio Astrofisico di Arcetri, Largo E. Fermi, 50125 Firenze, Italy \and Department of Astronomy, University of Maryland, College Park, USA}

\offprints{H. Beuther,\email{beuther@mpifr-bonn.mpg.de}}

\date{Received ... /Accepted ...}

%%%%%%%%%%%%%%%%%%%%%%%%%%%%%%%%%%%%%%%%%%%%%%%%%%%%%%%%%%%%%%%%%%%%%%%%%
%              abstract
%%%%%%%%%%%%%%%%%%%%%%%%%%%%%%%%%%%%%%%%%%%%%%%%%%%%%%%%%%%%%%%%%%%%%%%%%

\abstract{With the aim of understanding the role of massive outflows 
in high-mass star formation, we mapped in the $^{12}$CO $J=2-1$
transition 26 high-mass star-forming regions at very early stages of
their evolution. At a spatial resolution of $11''$ bipolar molecular outflows
were found in 21 of them. The other five sources show confusing
morphology but strong line wings. This high detection rate of bipolar
structure proves that outflows common in low-mass sources are also
ubiquitous phenomena in the formation process of massive stars. The
flows are large, very massive and energetic, and the data indicate
stronger collimation than previously thought. The dynamical timescales
of the flows correspond well to the free-fall timescales of the
associated cores. Comparing with correlations known for low-mass
flows, we find continuity up to the high-mass regime suggesting
similar flow-formation scenarios for all masses and
luminosities. Accretion rate estimates in the $10^4$~L$_{\odot}$ range
are around $10^{-4}$~M$_{\odot}$yr$^{-1}$, higher than required for
low-mass star formation, but consistent with high-mass star formation
scenarios. Additionally, we find a tight correlation between the
outflow mass and the core mass over many orders of magnitude. The
strong correlation between those two quantities implies that the
product of the accretion efficiency
$f_{\rm{acc}}=\dot{M}_{\rm{acc}}/(M_{\rm{core}}/t_{\rm{ff}})$ and
$f_{\rm{r}}$ (the ratio between jet mass loss rate and accretion
rate), which equals the ratio between jet and core mass
($f_{\rm{acc}} f_{\rm{r}}= M_{\rm{jet}}/M_{\rm{core}}$), is roughly
constant for all core masses. This again indicates that the
flow-formation processes are similar over a large range of
masses. Additionally, we estimate median $f_{\rm{r}}$ and
$f_{\rm{acc}}$ values of approximately 0.2 and 0.01, respectively,
which is consistent with current jet-entrainment models.\\ To
summarize, the analysis of the bipolar outflow data strongly supports
theories which explain massive star formation by scaled up,
but otherwise similar physical processes --~mainly accretion~-- to
their low-mass counterparts.
}

\maketitle

%%%%%%%%%%%%%%%%%%%%%%%%%%%%%%%%%%%%%%%%%%%%%%%%%%%%%%%%%%%%%%%%%%%%%%%%%%%%
%                  Introduction
%%%%%%%%%%%%%%%%%%%%%%%%%%%%%%%%%%%%%%%%%%%%%%%%%%%%%%%%%%%%%%%%%%%%%%%%%%%%

\section{Introduction}
\label{intro}

Molecular outflows are a well known phenomenon in sites of low-mass
star formation, and they play an important role in transporting
angular momentum away from the forming star. The observational
database on low-mass outflows has increased tremendously over the last
decade, giving rise to different formation scenarios. While it is
still unclear how the outflow is accelerated near the proto-star and/or
disk, it is now widely believed that low-mass molecular flows are
momentum driven by highly collimated jets, which entrain the
surrounding molecular gas. For recent reviews on this topic see,
e.g., \citet{richer 2000,bachiller 2000,shu 2000,koenigl 2000}.

The situation in the high-mass star formation regime is less clear,
because, due to the rarity of these objects and the typically larger
distances (a few kpc) of sites of high-mass star formation, the
spatial resolution has been lacking to resolve the outflows and their
driving sources properly. Recent systematic molecular line studies of
high-mass outflows have been carried out by \citet{shepherd 1996a,
shepherd 1996b, henning 2000, zhang 2001, ridge 2001}. All of them
find that $\sim 90\%$ of the observed massive star-forming regions are
associated with high velocity gas. Mapping a subsample of 10 sources
(out of 94)
\cite{shepherd 1996b} find bipolar morphology in 5 of
them and \cite{zhang 2001} found spatial outflow structures in 39 of
69 sources. Thus, both studies indicate bipolar structures in
at least $50\%$ of the observed sources.

Including more massive outflow sources from the literature, the
overall picture has emerged, that outflows are ubiquitous phenomena in
massive star formation, that they are very massive (up to hundreds of
solar masses in the flows), very energetic ($\sim 10^{46}$~erg) but
seemingly not very collimated (i.e. collimation factors --~the length
of the flow divided by its width~-- between 1 and 10 for low-mass
sources versus collimation factors between 1 and 1.8 for high-mass
sources, \citealt{richer 2000,churchwell 2000b,ridge 2001}). The high
masses in the flows as well as the low collimation factors, which are
difficult to explain in current jet-entrainment scenarios, have given
rise to new ideas of outflow formation including the possibility that
massive outflows consist of accelerated gas which has been deflected
by the young accreting proto-star, rather than swept-up ambient
material \citep{churchwell 2000b}. \citet{devine 1999} propose a
slightly different scenario in which massive stars produce collimated
jets only in their earliest phases, and with increasing luminosity of
the central objects the jets may be replaced by wide-angle winds.

To test the proposed scenarios it is necessary to increase the number
of outflow sources using a homogenous sample with adequate angular
resolution. Therefore, we mapped 26 promising candidates from a larger
sample of massive star formation sites in the $J=2\to 1$ line of CO
with the 30~m telescope at a spatial resolution of $11''$. This work
is part of a long term project to search and investigate 69 massive
and rather isolated star formation regions in an evolutionary stage
prior to building significant ultracompact H{\sc ii} regions. An
introduction to the sample and first results are given in
\cite{sridha} and \citet{beuther 2002a}.

In the following, we present the observed dataset and give a detailed
analysis of the outflow characteristics. The derived parameters are
compared with those from low-mass as well as other high-mass outflows,
and we discuss the implications for star formation in such regions.

%%%%%%%%%%%%%%%%%%%%%%%%%%%%%%%%%%%%%%%%%%%%%%%%%%%%%%%%%%%%%%%%%%%%%%%%%%%%
%             Observation 
%%%%%%%%%%%%%%%%%%%%%%%%%%%%%%%%%%%%%%%%%%%%%%%%%%%%%%%%%%%%%%%%%%%%%%%%%%%%

\section{Observations}
 
\subsection{Source Selection}

The 26 sources observed in this study were chosen from a sample of 69
high-mass proto-stellar candidates based on the CO 2--1 line wings
seen by these authors in pointed observations \citep{sridha}.  We do
not believe that there was a particular bias in this
selection. \citet{sridha} found wings in 85 percent of their sample
and argue on the basis of the statistical distribution of inclination
angles that essentially all of these sources are associated with
outflows.  As also argued by
\citet{sridha}, the majority of this sample is likely to be young and
50 percent have no associated 9 GHz radio continuum flux down to a
limit of 1 mJy. This suggests they have not had time to form a
substantial H{\sc ii} region.  While the low radio continuum flux may
have a variety of causes (e.g., high 9 GHz optical depth or dust mixed
with ionized gas), we consider the sample of sources observed by us to
be an arbitrarily chosen set of very young regions of high-mass star
formation.

\subsection{IRAM 30~m observations}

The IRAM 30~m telescope on Pico Veleta near Granada (Spain) was used
to map 26 sources in the $J=2-1$ transition of $^{12}$CO in April and
November 1999 and in April 2000. Maps with an average extent of $100''
\times 100''$ were observed in the on-the-fly mode, where the
telescope moves continously across the source and dumps the data in a
well chosen grid. At 230.5~GHz the resolution of the 30~m telescope is
approximately $11''$. \cite{beuther 2000} have shown that
Nyquist-sampling is sufficient in the on-the-fly mode as well, because
the beam is smeared out by only about $4\%$. Therefore we chose as a
sampling interval $4''$ (Nyquist interval $\theta
_{Ny}=\frac{\lambda}{2D}\sim 4.5''$). The dump time was 2 sec per
position and we scanned all maps twice, in most cases in perpendicular
directions to reduce scanning effects. Experience showed that the
whole system (weather+technical system, $T_{\rm{sys}}\sim 250$~K) is
stable for around 5 minutes. Therefore our ON-OFF cycles
never exceeded that time. OFF positions were chosen based on the
Stony Brook and the CfA-survey of the galactic plane \citep{sanders
1986,dame 2001} and checked to be emission free. The frequency
resolution was 0.1~km/s and the beam efficiency 0.41.

The data were reduced with CLASS and GRAPHIC of the GILDAS software
package by IRAM and the Observatoire de Grenoble. To improve the
signal to noise ratio, the $^{12}$CO data are smoothed to a velocity
resolution of 1~km/s, sufficient to sample the broad CO lines.

\begin{figure*}[h]
\includegraphics[angle=-90,width=15cm]{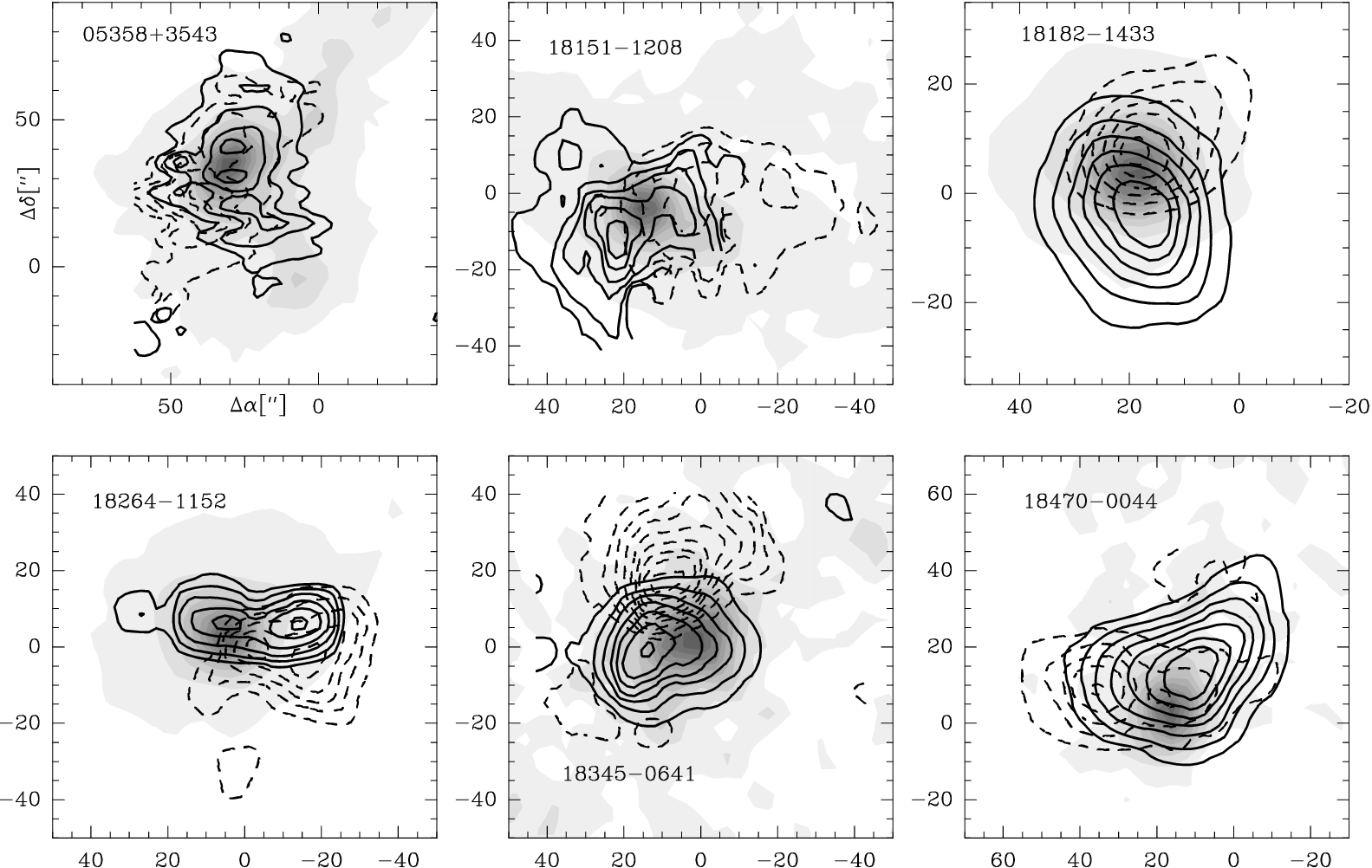}
\includegraphics[angle=-90,width=15cm]{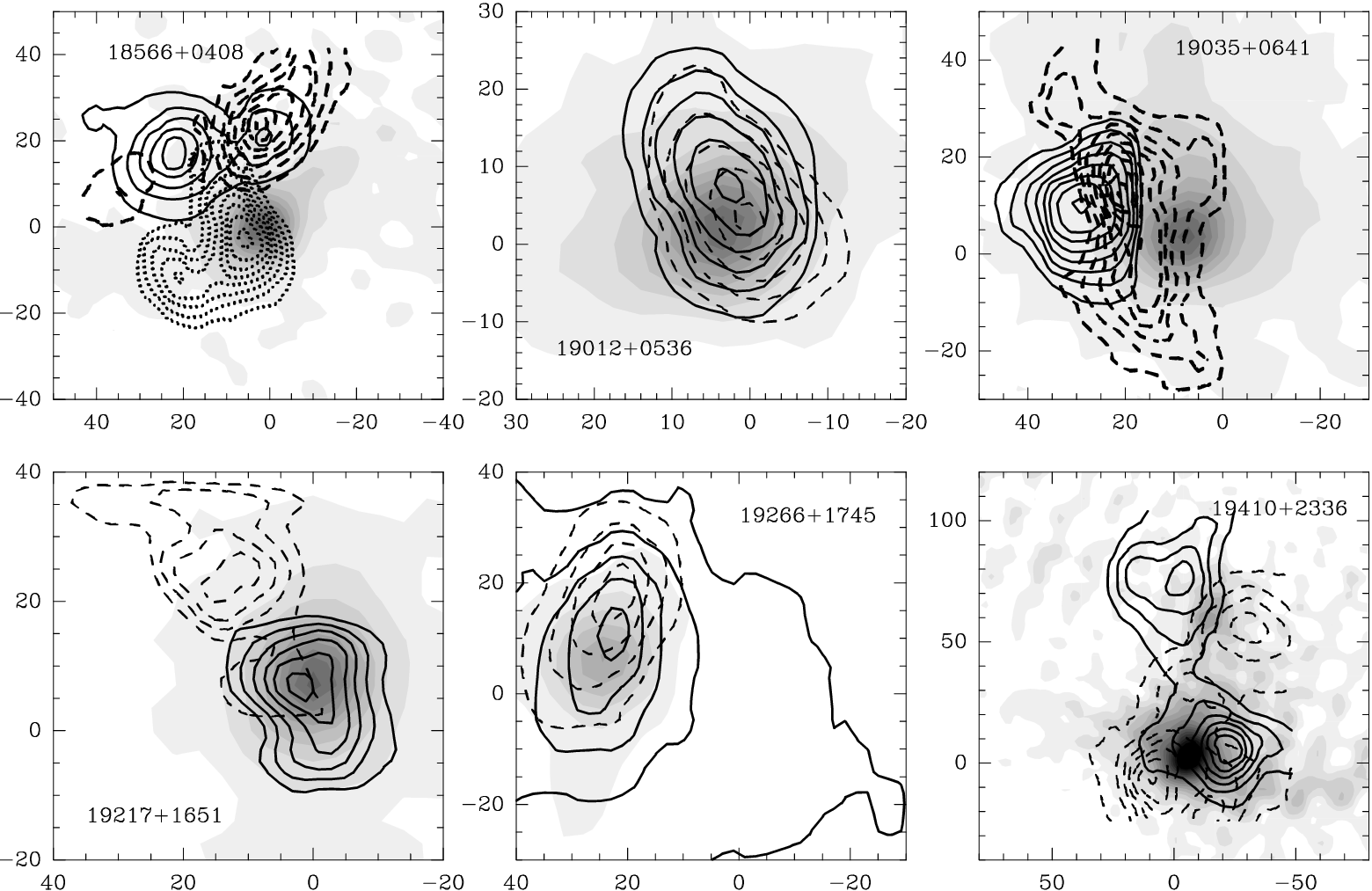}
\caption{Massive molecular outflows: the solid lines show the blue wings and the dashed lines the red wings of CO 2--1 emission. The contours are chosen to highlight the most prominent features in each source, usually between $30\%$ and $90\%$ (steps of $10\%$) of the peak integrated wing intensity. The dotted lines in 18566+0408 represent an additional feature at ``blue'' velocities [58,60]~km/s. The grey scale presents the corresponding 1.2~mm continuum maps from $5\%$ and $95\%$ by steps of $10\%$ of the peak flux as outlined in \cite{beuther 2002a}. The axis show offsets in arcsec from the absolute {\it IRAS}-positions given in \citet{sridha}. Detailed morphological descriptions are given in Sect.\ref{morphologies} \label{outflows}.}
\end{figure*}

\setcounter{figure}{0}

\begin{figure*}[ht]
\includegraphics[angle=-90,width=15cm]{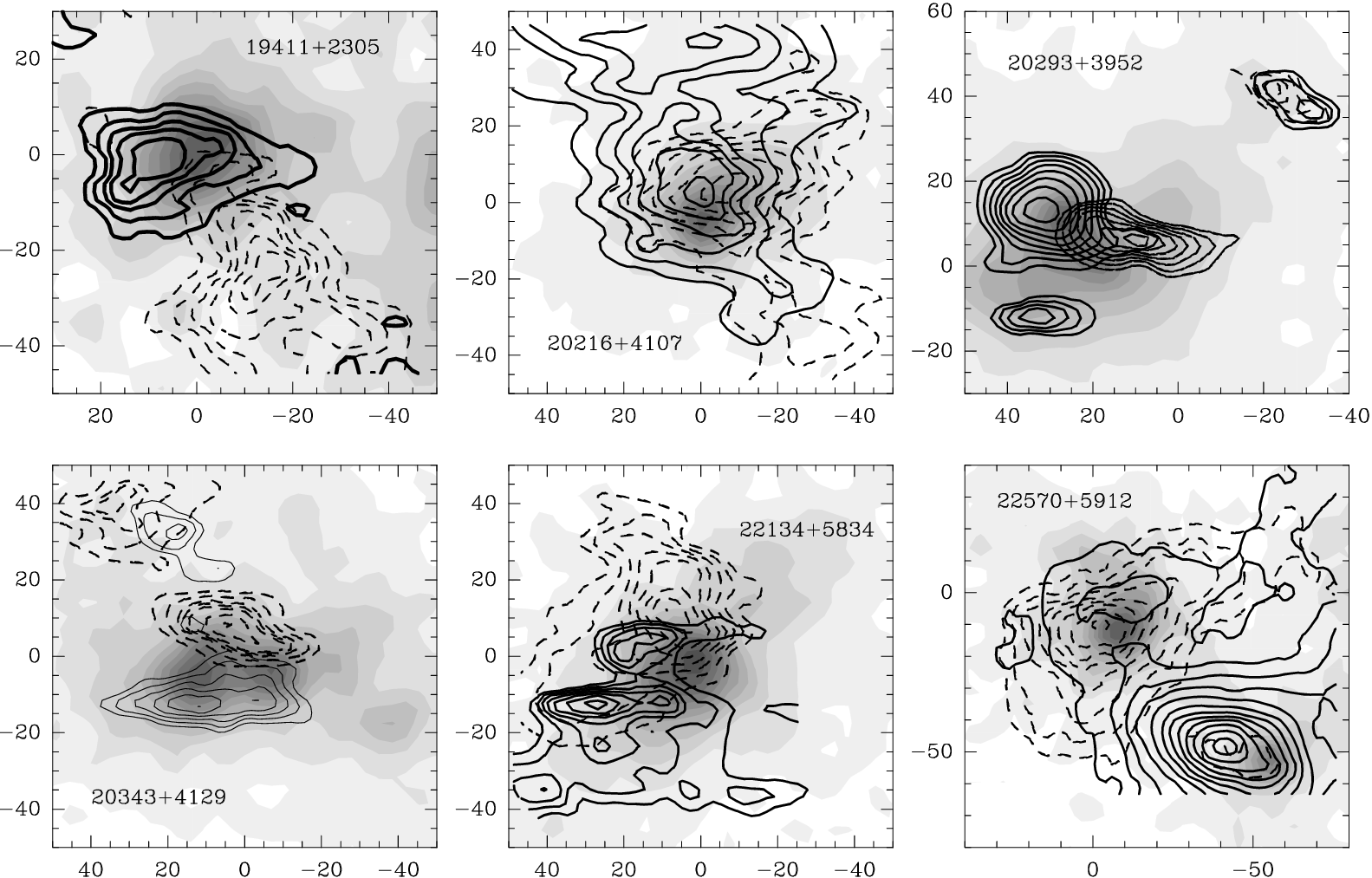}
\includegraphics[angle=-90,width=15cm]{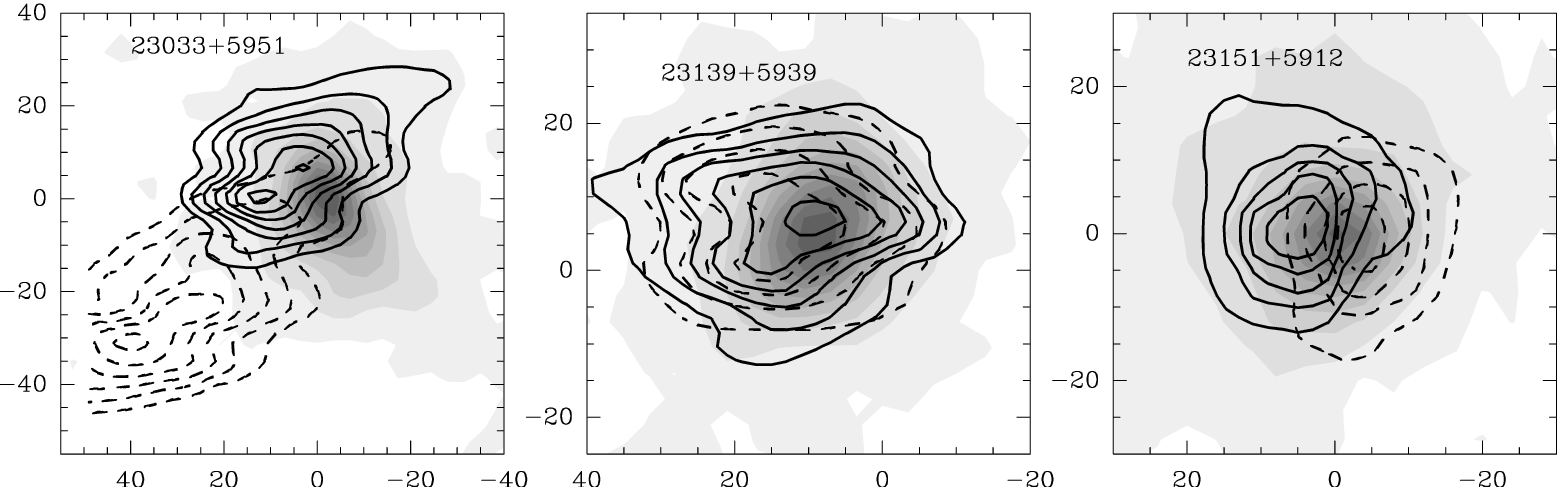}
\caption{continued. The red wing for the southern flow in 20293+3952 is missing due to instrumental problems in that region, but simultaneously observed SiO data (not presented here) confirm this southern flow.}
\end{figure*}

\begin{figure*}[ht]
\includegraphics[angle=-90,width=18cm]{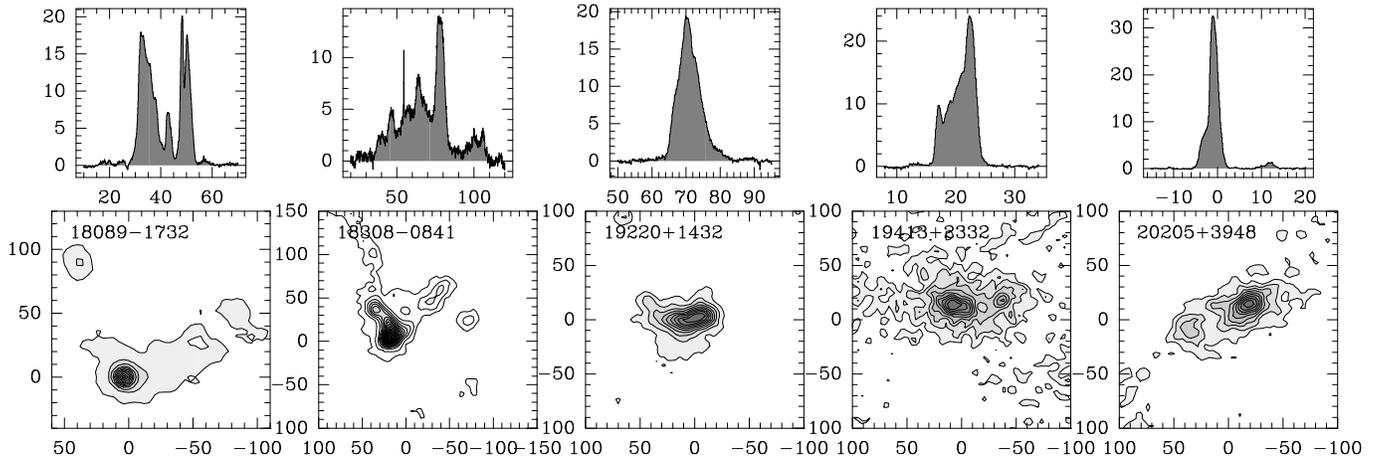}
\caption{For sources without clear bipolar outflow structure we show the summed $^{12}$CO 2--1 spectra ($\rm{T_{mb}}$ [K] versus velocity [km/s]) corresponding to the regions outlined by the 1.2~mm continuum maps in grey scale. The continuum images are presented as in Fig. \ref{outflows} \label{nooutflows}.}
\end{figure*}

\clearpage

%%%%%%%%%%%%%%%%%%%%%%%%%%%%%%%%%%%%%%%%%%%%%%%%%%%%%%%%%%%%%%%%%%%%%%%%%%%%
%          Observational results
%%%%%%%%%%%%%%%%%%%%%%%%%%%%%%%%%%%%%%%%%%%%%%%%%%%%%%%%%%%%%%%%%%%%%%%%%%%%

\section{Observational results}

\subsection{Morphologies}
\label{morphologies}

Fig. \ref{outflows} presents the outflow sources [CO 2--1 blue (solid
lines) and red wing emission (dashed lines) overlayed on the grey
scale 1.2~mm continuum maps \citep{beuther 2002a}]. From the 26 sources
observed, 21 show either bipolar morphology or wide non--Gaussian
line wings such that we suspect the outflow is along the line of
sight (23139+5939) and not spatially resolvable with the $11''$
resolution of our observations. The 5 remaining sources ($18089-1732$,
$18308-0841$, $19220+1432$, $19413+2332$ and $20205+3948$) also show
line wings, but the maps are very chaotic, so that we could not define
any outflow structure. All the sources are in the galactic plane
\citep{sridha} and unrelated spiral arm emission is  partly
confusing the spectra. Fig. \ref{nooutflows} presents for  ``confused''
sources the sum of all spectra in the regions outlined by the 1.2~mm
dust maps \citep{beuther 2002a}. We cannot tell if the latter 
are a superposition of many flows, or if they are
more isotropic flows, similar to that observed in Orion
\citep{rodriguez 1999,schultz 1999}. This can only be investigated
with higher angular resolution in the future but we note that most of
our sample is less luminous than Orion IRC2 ($L\sim 2\times
10^5~\rm{L}_{\odot}$, \citealt{churchwell 2000b}). Within the range of
luminosities which we sample (roughly $10^{3.5}$ to
$10^5$~L$_{\odot}$), we find that bipolar features are omnipresent.
We also do not observe a statistical difference in finding
bipolar outflow structures for sources with or without detected cm
emission suggesting perhaps that the arrival of proto-stars
on the ZAMS is not critical for the outflow.

All the mapped sources showed wing emission in previous single
pointing observations conducted at the 30~m telescope
\citep{sridha}. Therefore we expected outflows in all sources, but the
large fraction with bipolar structures ($\sim 80\%$) is rather
surprising. Previous mapping studies of massive molecular outflows found
bipolar structures only in approximately $50\%$ of the sources
\citep{shepherd 1996a,zhang 2001}, which seemingly suggested
that bipolarity is less probable in high-mass than in low-mass star formation
\citep{churchwell 2000b}. The high detection rate of bipolar outflows 
in our study is most likely due to the improved spatial resolution of
our data ($11''$) compared to previous statistical work on massive
outflows (compare radius $r$ in Table \ref{input} to $60''$ resolution
in \citet{shepherd 1996b} and $30''$ in \citet{zhang 2001}).

Every outflow is associated with massive mm cores (grey-scale in Figure
\ref{outflows}). It is not evident that the size of
the flows depends on the size of the cores: about $50\%$ of the flows
have an extent larger than the mapped core sizes, whereas the other
half of the flows are smaller. In most cases, the flows are centered
on the mm peak, which most likely harbors the most massive and
youngest proto-stars. But there are a few remarkable exceptions, where
the flows are centered offset from the mm peak (18566+0408,
19035+0641, 19217+1651, 19411+2305, 22570+5912). The reasons for such
offsets are not clear. For 18566+0408 and 19035+0641, pointing
problems may be the explanation since simultaneously observed
H$^{13}$CO$^+$~1--0 maps (not published so far) peak approximately at
the center of the flows and offset relative to the mm cores. In the
cases of the more chaotic looking flows 19411+2305 and 22570+5912,
this seems unlikely and the structure may be due to several low-mass
flows, which are not at the center of the mm core.

19217+1651 on the other hand is not chaotic but a nice examples of a
bipolar outflow which makes a superposition of several low-mass flows
very unlikely. Additionally, the offset between the flow center and
the single dish outflow map is real because H$^{13}$CO$^+$~1--0 as
well as thermal CH$_3$OH (both observed simultaneously with the
outflow) peak at the mm core. Using a theoretically
derived correlation between the momentum of the outflow and the
proto-stellar mass (Eq. 18 in
\citet{tan 2002}), we can infer a mass for the proto-star powering the
outflow. In this way, we conclude that the proto-star has a mass of at
least 10~M$_{\odot}$.  This suggests that high-mass proto-stars can
emerge from the core where they were formed. The projected offset
between the flow center and the mm peak is of order 0.3 pc which
implies a relatively high peculiar velocity of the proto-star relative
to the core ($\sim 4$~km/s) using the flow timescale given in Table
\ref{output}.  This is obviously an object of considerable interest
and interferometric observations with the Plateau de Bure
Interferometer are currently being conducted.

The flows are rather large with an average size of $\sim 0.8$~pc (see
Table \ref{output}). The collimation factors derived for this sample
are higher than previously claimed. \cite{richer 2000} in their review
note that so far no massive flow with a collimation factor
$f_{\rm{c}}$ larger than 1.8 has been observed. The last column of
Table \ref{input} presents the collimation factors we derive for our
sample (flows with a question mark are along the line of sight, which
makes a determination of $f_{\rm{c}}$ difficult). These collimation
factors are lower limits due to angular resolution and projection
effects, but we nevertheless obtain a mean $f_{\rm{c}}$ of 2.1, larger
than most measurements of high-mass flows to date.  The largest
$f_{\rm{c}}$ we find in the single dish data is 3, but we know from
interferometric observations of selected sources that $f_{\rm{c}}$ can
be even higher: $f_{\rm{c}}\sim 10$ for 05358+3543 \citep{beuther
2002b} and $f_{\rm{c}}\sim 5$ for 23033+5951 (Wyrowski et al., in
prep.). It is interesting to compare the observed collimation factors
with theoretical expectations: a flow at 5~kpc (the
mean distance of this sample) extended 1~pc in length and 0.1~pc in
width results in an intrinsic $f_{\rm{c}}=10$. Inclination effects to
first order just reduce the length of the flow but not the widths;
thus the length has to be corrected on the average by
$\cos(57^{\circ})$ (mean inclination angle). After converting the
linear scales into arcseconds and convolving with an $11''$ beam
--~as for our CO data ~-- the length of the flow is $\sim 25''$ and
the width $\sim 12''$, resulting in an observable $f_{\rm{c}}=2.1$,
which is similar to the mean value we derive for our sample. Thus, the
observed collimation factor is a strong function of the available
angular resolution and high-mass flows may in general be as well
collimated as flows from low-mass proto-stars (though see the
discussions of \citet{devine 1999} and \citet{reipurth
2001}). Even this new dataset with $11''$ resolution underestimates
collimation factors; higher resolution with interferometers is needed
for that.

In 4 fields (18151$-$1208, 19410+2336, 20293+3952, 20343+4129) we see
at least two separated flows. Interferometric data for two of the
flows (05358+3543 \citealt{beuther 2002b}; 23033+5951 Wyrowski et
al., in prep.) reveal that those single dish flows split up into at
least two outflows, but the data also show that most of the
single dish flux ($>50\%$ for 05358+3543 and $>80\%$ for 23033+5951)
is produced by one major flow with minor contributions from the second
flow. There are likely to be multiple lower-mass flows, and recent
theoretical studies by \citet{tan 2002} consider massive outflows to
be a superposition of outflows from all proto-stars of each cluster
evolving simultaneously. This scenario results in less collimated
flows and may explain some of our more chaotic sources, but it does
not seem to be applicable in general. Thus, contributions from
lower-mass flows are possible, but the high degree of collimation and
bipolarity we find on average in addition to that found in a small
number of sources with interferometric observations strongly suggest
that our data are dominated by the most massive outflow of each
cluster.

\subsection{Characteristics of the outflows}

The determination of outflow parameters is subject to a large number
of errors. To first order, it is difficult to separate the outflowing
gas  from the ambient gas, and the inclination angles of the
flows are also  often unknown. \cite{cabrit 1986, cabrit 1990, cabrit
1992} performed a series of studies on outflows from low-mass sources
and the approach we are following here is outlined in
\cite{cabrit 1990}. The velocity range due to the outflow is
determined in two ways. One is to define the flow by the line wings in
the spectra, the other approach is to map each channel and decide from
the spatial separation of different channels, which belong to an
outflow and which correspond to the ambient core emission. In some sources
both criteria were suitable, in other sources only one of them,
e.g., the 05358+3543 flow is known to be near the plane of the sky
\citep{beuther 2002b} causing us to use the spatial separation. A
counter example is 23139+5939, which shows extremely strong wing
emission but nearly no lobe separation probably due to being rather
along the line of sight (Fig. \ref{outflows}). Then we mapped the
chosen wings, and the size of the flow (transformed to linear scales,
size$_{\rm{b}}$ \& size$_{\rm{r}}$) and the mean value of integrated
wing emission ($\int T_{\rm{mb}}(^{12}\rm{CO} 2-1)$, mean$_{\rm{b}}$
\& mean$_{\rm{r}}$) were used as inputs to the flow
calculations. Additional input parameters are the radius $r$ of the
flow from the projected center, and the maximum velocity separation of
the red and the blue wings $v_{\rm{max_r}}$
\& $v_{\rm{max_b}}$. All the input parameters are presented in Table
\ref{input}. An additional uncertainty is caused by distance 
ambiguity in which cases the outflow parameters are
calculated for the near and the far distance (Table \ref{output}).

\begin{table*}
\caption{Parameters for outflow calculations: The number of flows per source \#, the infrared luminosity $L$ derived from the {\it HIRES}-data \citep{sridha} (if the kinematic ambiguity is unresolved both values --~near and far~-- are given), the velocity range of the wings $\Delta v$, the mean of the integrated wing emission, the maximum projected velocity $v_{\rm{max}}$ and the radius $r$ of the flow. The last column presents the collimation factor $f_{\rm{c}}$ \label{input}}
\begin{tabular}{lcrrrrrrrrr}
source & \# & $L$ & $\Delta v_{\rm{b}}$ & $\Delta v_{\rm{r}}$ & mean$_{\rm{b}}$ & mean$_{\rm{r}}$ &
$v_{\rm{max_b}}$ & $v_{\rm{max_r}}$ & $r$ & $f_{\rm{c}}$ \\ & & [log(L$_{\odot}$)] & [km/s] & [km/s] &
[km/s K] & [km/s K] & [km/s] & [km/s] & [$''$] & \\
\hline
05358+3543 &  1 & 3.8 & ($-$32,$-$21)   & ($-$12,$-$4)  & 28.5 & 29.7  &   14.4 &  13.6  &   60 & 2.0 \\ 
18151$-$1208 &  1 & 4.3 & (25,30) & (38,42) & 20.2 & 12.2              &   5    &  10    &   45 & 2.1 \\
18182$-$1433 &  1 &  4.3/5.1 & (53,56) & (62,70) & 26.5 & 26.1              &   6.1  &  12    &   25 & 1.7 \\
18264$-$1152 &  1 &  4.0/5.1 & (28,37) & (52,63) & 38.4 & 46.6              &   16   &  18    &   20 & ? \\
18345$-$0641 &  1 &  4.6 & (82,90) & (102,106) & 28.6 & 16.8            &   14   &  10    &   40 & 1.5 \\
18470$-$0044 &  1 &  4.9 & (86,91) & (102,106) & 24.3 & 9.0             &   10.5 &  9.5   &   38 & 2.0 \\
18566+0408  &  1 &  4.8 & (68,77) & (96,102) & 15.2 & 9.2               &   17   &  17    &   40 & 2.4 \\
19012+0536  &  1 &  4.2/4.7 & (55,62) & (70,75) & 39.6 & 21.6               &   10   &  10    &   30 & ? \\
19035+0641  &  1 &  3.9 & (21,27) & (38,43) & 12.0 & 4.2                &   11   &  11    &   30 & ? \\
19217+1651 &  1 &  4.9 & ($-$16,$-$3) & (9,21)    & 31.4 & 26.2         &   19.5 &  17.5  &   28 & 2.8 \\ 
19266+1745 &  1 &  1.7/4.7 & ($-3,0$) & (10,15) & 15.4 & 5.4                &   8    &  10    &   35 & ? \\
19410+2336 &  1 &  4.0/5.0 & (1,17) & (26,45) & 35.3 & 64.0                 &   21.4 &  22.5  &   43 & 2.0\\
19410+2336 &  2 &  4.0/5.0 & (1,17) & (26,45) & 36.0 & 52.0                 &   21.4 &  22.5  &   45 & 2.1 \\
19411+2306 &  1 &  3.7/4.3 & (15,25) & (35,43) & 32.4 & 27.0                &   14   &  14    &   35 & 2.3 \\
20216+4107 &  1 &  3.3 & ($-$9,$-$4) & (1,6) & 30.8 & 28.8             &   6.6  &  8.2   &   32 & ? \\ 
20293+3952 &  1 &  3.4/3.8 & ($-$16,4) & (12,25)   & 25.6 & 61.6            &   26   &  30    &   32 & 2.7 \\
20293+3952 &  2 &  3.4/3.8 & ($-$16,4) & (12,25)   & 18.2 & 3.4             &   26   &  30    &   15 & ? \\
20343+4129 &  1 &  3.5 & (8,9)       & (13,15) & 24.6 & 33.4            &   3.5  &  4.5   &   23 & 1.0 \\
20343+4129 &  2 &  3.5 & (8,9)       & (13,15) & 9.2 & 12.0             &   3.5  &  4.5   &   25 & 1.7 \\
22134+5834 &  1 &  4.1 & ($-$30,$-$23)   & ($-$14,$-$4)    & 10.6 & 34.4&   11.7 &  14.3  &   37 & ? \\
22570+5912 &  1 &  4.7 & ($-56,-50$) & ($-42,-32$) & 20.8 & 17.8        &   10   &  14    &   60 & ? \\
23033+5951 &  1 &  4.0 & ($-$67,$-$60   )& ($-$47,$-$33)   & 21.8 & 34.2&   14.0 &  19.4  &   50 & 3.0 \\
23139+5939 &  1 &  4.4 & ($-$59,$-$47)   & ($-$39,$-$31)   & 43.0 & 23.6&   11.1 &  13.2  &   29 & ? \\
23151+5912 &  1 &  5.0 & ($-89,-65$) & ($-47,-35$) & 53.8 & 11.8        &   34   &  19    &   20 & 1.2 \\
\end{tabular}
\end{table*}

Opacity corrected H$_2$ column densities $N_{\rm{b}}$ and $N_{\rm{r}}$
in both outflow lobes can be calculated by assuming a constant
$^{13}$CO/$^{12}$CO 2--1 line wing ratio throughout the outflows
\citep{cabrit 1990}. \citet{choi 1993} found an average
$^{13}$CO/$^{12}$CO 2--1 line wing ratio around 0.1 in 7 massive
star-forming regions, which we adopt for our sample as
well. \citet{levreault 1988} found similar values in low-mass
outflows. The derived flow parameters are the masses $M_{\rm{b}}$,
$M_{\rm{r}}$ in the blue and red outflow lobes and the total mass
$M_{\rm{out}}$, the momentum $p$, the energy $E$, the size, the
characteristical time scale $t$ (radius of the flow $r$ divided by the
flow velocity), the mass entrainment rate of the molecular outflow
$\dot{M}_{\rm{out}}$ (as opposed to the mass loss rate of the jet
$\dot{M}_{\rm{jet}}$), the mechanical force $F_{\rm{m}}$ and the
mechanical luminosity $L_{\rm{m}}$ for each flow. For details see
\cite{cabrit 1990}. The used Eqs. are (assuming 
$\rm{H}_2/^{13}\rm{CO}=89\times 10^4$ \citep{cabrit 1992} and
$T_{\rm{ex}}=30$~K; $m_{\rm{H_2}}$ is the mass of the H$_2$ molecule):
\begin{eqnarray}
N &=& \left( \frac{\rm{H}_2}{^{13}\rm{CO}} \right) \left(\frac{\int T_{\rm{mb}}(^{13}\rm{CO} 2-1)}{\int T_{\rm{mb}}(^{12}\rm{CO} 2-1)} \right) \frac{3k^2T_{\rm{ex}}}{4\pi^3h\nu^2\mu^2} \\
  &  & \times ~ e^{-16.6/T_{\rm{ex}}} \int T_{\rm{mb}}(^{12}\rm{CO} 2-1)\\
M_{\rm{out}} &=& (N_{\rm{b}} \times \rm{size_b} + N_{\rm{r}} \times \rm{size_r})\ m_{\rm{H_2}} \\
p &=& M_{\rm{b}}\times v_{\rm{max}_b} + M_{\rm{r}}\times v_{\rm{max}_r} \\
E &=& \frac{1}{2}M_{\rm{b}}\times v_{\rm{max}_b}^2 + 
      \frac{1}{2}M_{\rm{r}}\times v_{\rm{max}_r}^2 \\
t &=& \frac{r}{(v_{\rm{max}_b}+v_{\rm{max}_r})/2}\\
\dot{M}_{\rm{out}} &=& \frac{M_{\rm{out}}}{t} \\
F_{\rm{m}} &=& \frac{p}{t}\\
L_{\rm{m}} &=& \frac{E}{t} 
\end{eqnarray} 
In the case of 18151$-$1208, we just determined the characteristics for
the eastern flow, because the western one is confused by ambient
emission. The derived quantities are shown in Table \ref{output}.
\cite{cabrit 1990} state the mass
determinations to be correct within approximately a factor 2, while
kinematic parameters should be approximately correct within a factor
10. It is clear however from the \citet{choi 1993} study that the high
opacity in $^{12}$CO transitions can lead to large errors for individual
objects and it would be useful to obtain $^{13}$CO data for these
sources. While several of the derived parameters are discussed in detail in \S
\ref{discussion}, we want to stress that all of the observed sources 
show massive and energetic molecular outflows compared to
low-mass flows \citep{richer 2000}.
 
\begin{table*}
\caption{Outflow results: distances $D$ [kpc], {\bf number of outflows \#, column densities $N$ [log$_{10}$(cm$^{-2}$)], masses $M_{\rm{b}}$ (blue), $M_{\rm{r}}$ (red) and $M_{\rm{out}}$ ($M_{\rm{out}}=M_{\rm{b}}+M_{\rm{r}}$) [$M_{\odot}$],} momentum $p$ [$M_{\odot}$ km/s], energy $E$ [$10^{46}$erg], size [pc], time $t$ [$10^4$yr], mass entrainment rate $\dot{M}_{\rm{out}}$ [$10^{-4}$M$_{\odot}$/yr], mechanical force $F_{\rm{m}}$ [$10^{-3}$M$_{\odot}$km/s/yr] and mechanical luminsoity $L_{\rm{m}}$ [$L_{\odot}$] \label{output}}
\begin{tabular}{lrrrrrrrrrrrrrr}
source & $D$ & \# & $N_{\rm{b}}$ & $N_{\rm{r}}$ & $M_{\rm{b}}$ & $M_{\rm{r}}$ & $M_{\rm{out}}$ & $p$ & $E$ & size & $t$ & $\dot{M}_{\rm{out}}$ & $F_{\rm{m}}$ & $L_{\rm{m}}$ \\ 
\hline
05358+3543 &  1.8 & 1 &  21.1 &  21.2 &   11 &   10 &   21 &   288 & 4.0 &  0.52 &  3.7 & 5.6 & 7.9 &   9.1 \\ 
18151-1208 &  3.0 & 1 &  21.0 &  20.8 &    8 &    4 &   12 &   106 & 0.9 &  0.65 &  7.1 & 1.7 & 1.5 &   1.1 \\ 
18182-1433 & 11.7 & 1 &  21.1 &  21.1 &  130 &   73 &  203 &  1665 & 15.0 &  1.42 & 15.3 & 13.0 & 11.0 &   8.2 \\ 
18182-1433 &  4.5 & 1 &  21.1 &  21.1 &   19 &   11 &   30 &   246 & 2.3 &  0.55 &  8.9 & 3.4 & 2.8 &   2.1 \\ 
18264-1152 & 12.5 & 1 &  21.3 &  21.4 &  234 &  171 &  405 &  6818 & 110 &  1.21 &  7.0 & 58.0 & 98.0 & 135.0 \\ 
18264-1152 &  3.5 & 1 &  21.3 &  21.4 &   18 &   13 &   31 &   534 & 9.0 &  0.34 &  3.7 & 8.6 & 14.0 &  20.0 \\ 
18345-0641 &  9.5 & 1 &  21.1 &  20.9 &  103 &   40 &  143 &  1841 & 24.0 &  1.84 & 15.0 & 9.5 & 12.0 &  13.1 \\ 
18470-0044 &  8.2 & 1 &  21.1 &  20.6 &   86 &   16 &  102 &  1050 & 11.0 &  1.51 & 14.8 & 6.9 & 7.1 &   6.0 \\ 
18566+0408 &  6.7 & 1 &  20.9 &  20.7 &   25 &    7 &   32 &   540 & 9.1 &  1.30 &  7.5 & 4.3 & 7.2 &  10.0 \\ 
19012+0536 &  8.6 & 1 &  21.3 &  21.0 &  105 &   29 &  134 &  1339 & 13.0 &  1.25 & 12.2 & 11.0 & 11.0 &   8.9 \\ 
19012+0536 &  4.6 & 1 &  21.3 &  21.0 &   30 &    8 &   38 &   383 & 3.8 &  0.67 & 13.0 & 2.9 & 2.9 &   2.4 \\ 
19035+0641 &  2.2 & 1 &  20.8 &  20.3 &    2 &    1 &    3 &    28 & 0.3 &  0.32 &  2.9 & 0.9 & 1.0 &   0.9 \\ 
19217+1651 & 10.5 & 1 &  21.2 &  21.1 &   38 &   70 &  108 &  1961 & 36.0 &  1.43 &  7.5 & 14.0 & 26.0 &  38.8 \\ 
19266+1745 & 10.0 & 1 &  20.9 &  20.4 &   17 &   18 &   35 &   311 & 2.8 &  1.70 & 18.4 & 1.9 & 1.7 &   1.3 \\ 
19266+1745 &  0.3 & 1 &  20.9 &  20.4 & 0.02 & 0.02 & 0.04 &   0.3 & 0.003 &  0.05 & 1.0 & 0.03 & 0.03 &   0.02 \\ 
19410+2336 &  6.4 & 1 &  21.2 &  21.5 &   92 & 1331 & 1423 & 31896 & 710 &  1.33 &  5.9 & 240 & 540 & 982.9 \\ 
19410+2336 &  2.1 & 1 &  21.2 &  21.5 &   10 &  143 &  153 &  3434 & 77.0 &  0.44 &  3.8 & 40.0 & 90.0 & 165.3 \\ 
19410+2336 &  6.4 & 2 &  21.2 &  21.4 &   96 &   77 &  173 &  3766 & 82.0 &  1.40 &  6.2 & 28.0 & 61.0 & 108.3 \\ 
19410+2336 &  2.1 & 2 &  21.2 &  21.4 &   10 &    8 &   18 &   405 & 8.8 &  0.46 &  4.0 & 4.6 & 10.0 &  18.2 \\ 
19411+2306 &  5.8 & 1 &  21.2 &  21.1 &   19 &   28 &   46 &   649 & 9.0 &  0.98 &  6.9 & 6.7 & 9.4 &  10.8 \\ 
19411+2306 &  2.9 & 1 &  21.2 &  21.1 &    5 &    7 &   12 &   162 & 2.3 &  0.49 &  6.9 & 1.7 & 2.4 &   2.7 \\ 
20216+4107 &  1.7 & 1 &  21.2 &  21.1 &    3 &    3 &    6 &    43 & 0.3 &  0.26 &  3.5 & 1.7 & 1.3 &   0.8 \\ 
20293+3952 &  2.0 & 1 &  21.1 &  21.5 &    2 &    7 &    9 &   270 & 7.8 &  0.31 &  1.1 & 8.6 & 25.0 &  59.2 \\ 
20293+3952 &  1.3 & 1 &  21.1 &  21.5 &    1 &    3 &    4 &   114 & 3.3 &  0.20 &  1.3 & 3.0 & 8.7 &  20.6 \\ 
20293+3952 &  2.0 & 2 &  20.9 &  20.2 &    1 &  0.1 &  1.1 &    21 & 0.6 &  0.15 &  0.5 & 1.6 & 4.2 &   9.1 \\ 
20293+3952 &  1.3 & 2 &  20.9 &  20.2 &  0.3 &  0.1 &  0.4 &     8 & 0.2 &  0.09 &  0.6 & 0.6 & 1.5 &   3.2 \\ 
20343+4129 &  1.4 & 1 &  21.1 &  21.2 &    1 &    1 &  2   &     9 & 0.04 &  0.16 &  3.8 & 0.6 & 0.2 &   0.1 \\ 
20343+4129 &  1.4 & 2 &  20.7 &  20.8 &  0.2 &  0.2 &  0.4 &     1 & 0.007 &  0.17 &  4.2 & 0.1 & 0.04 &   0.01 \\ 
22134+5834 &  2.6 & 1 &  20.7 &  21.2 &    2 &   15 &   17 &   242 & 3.4 &  0.47 &  3.5 & 4.9 & 6.9 &   7.9 \\ 
22570+5912 &  5.1 & 1 &  21.0 &  20.9 &   30 &   43 &   73 &   911 & 11.0 &  1.48 & 12.1 & 6.1 & 7.5 &   7.8 \\ 
23033+5951 &  3.5 & 1 &  21.0 &  21.2 &    9 &   23 &   32 &   566 & 10.0 &  0.85 &  5.0 & 6.4 & 11.0 &  16.9 \\ 
23139+5939 &  4.8 & 1 &  21.3 &  21.1 &   41 &   16 &   57 &   662 & 7.7 &  0.67 &  5.4 & 10.0 & 12.0 &  11.7 \\ 
23151+5912 &  5.7 & 1 &  21.4 &  20.8 &   13 &    8 &   21 &   597 & 18.0 &  0.55 &  2.0 & 10.0 & 29.0 &  72.7 \\ 
\end{tabular}
\end{table*}

\begin{figure}[ht]
\includegraphics[angle=-90,width=8cm]{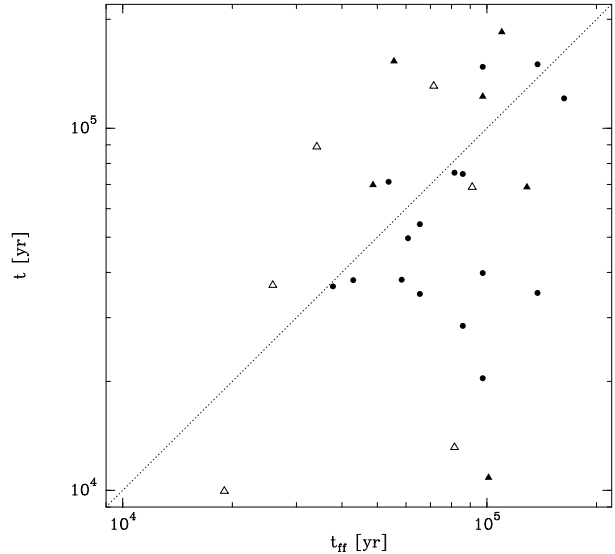}
\caption{Dynamical outflow times $t$ versus free-fall times $t_{\rm{ff}}$.
The circles show sources with known distance whereas triangles represent sources 
with unresolved distance ambiguity (open triangle: near distance, filled triangle: far distance). \label{tt}}
\end{figure}

Another important result from our study concerns the derived flow
timescales $t$. These are likely to represent a lower limit to the true
age of the flows (see \citealt{parker 1991}) and hence also to the
time over which the embedded proto-stars responsible for the flows have
been accreting from their surroundings.  One can compare these
estimates for the outflows with the free-fall times ($t_{\rm{ff}}=
\sqrt{\frac{3\pi}{32G\rho}}$) derived from the mean gas density $\rho $ 
in the cores from which presumably the accretion takes place.  $\rho $
is derived from the dust continuum maps as discussed by \cite{beuther
2002a}.  The free-fall times are an estimate of the timescale for
dynamical evolution of the core and in fact theoretical studies of
core evolution suggest that the timescale for star formation in a
core is of order a few free-fall times (e.g., \citealt{tan 2002}).
Our comparison of $t$ and
$t_{\rm{ff}}$ is shown in Fig. \ref{tt}. We see that flow timescales and
free-fall timescales are of the same order of magnitude and range for
our sample between $2\times 10^4$ and $2\times 10^5$ years with an average
dispersion about a factor 2 (for the sources with known distance). The
rough equality between these independently derived quantities is
surprising and it will be important to check whether this also holds for
other samples.  We tentatively conclude that for the present sample,
flow ages are good estimates of proto-star lifetimes and will assume
this in the following discussion.

%%%%%%%%%%%%%%%%%%%%%%%%%%%%%%%%%%%%%%%%%%%%%%%%%%%%%%%%%%%%%%%%%%%%%%%%%%
%                            Discussion
%%%%%%%%%%%%%%%%%%%%%%%%%%%%%%%%%%%%%%%%%%%%%%%%%%%%%%%%%%%%%%%%%%%%%%%%%%

\section{Discussion} 
\label{discussion} 

\subsection{High-mass versus low-mass outflows}
\label{high_low}

We wish first of all to compare our results to systematic studies
conducted in the low-mass regime. The analysis used for the low-mass
flows discussed by \cite{cabrit 1992} and \citet{bontemps 1996} is
very similar to ours and is based on results of \cite{cabrit
1990}. Hence, we can usefully compare our data with those results.

\begin{figure*}[ht]
\includegraphics[angle=-90,width=16cm]{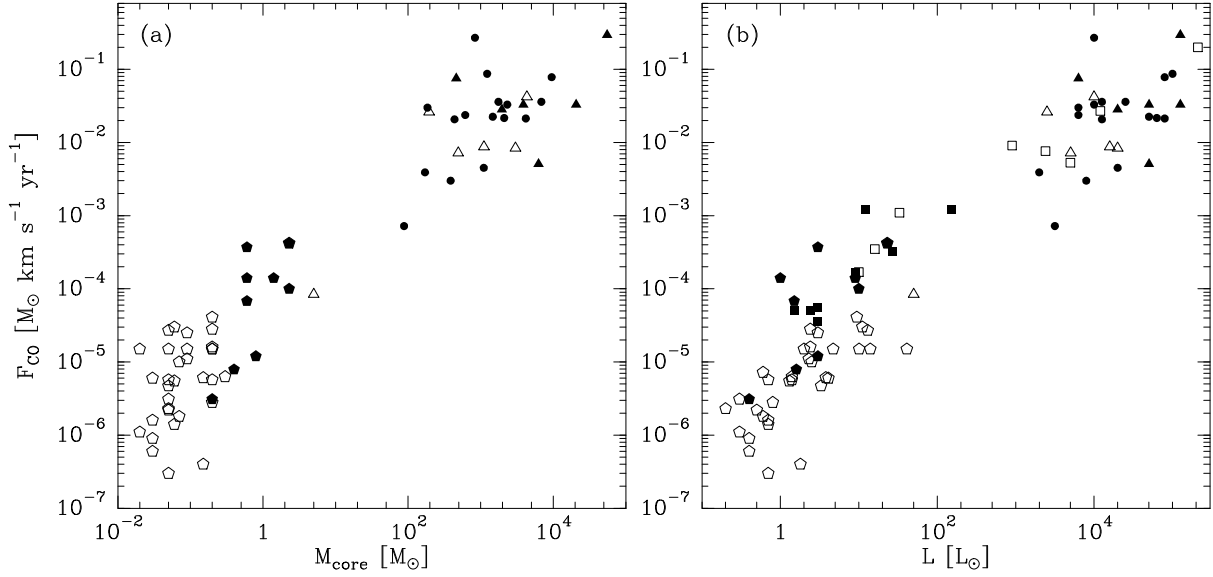}
\caption{Outflow parameter correlations: {\bf (a)} mechanical force from CO versus core mass from 1.2~mm dust emission. {\bf (b)} mechanical force versus bolometric luminosity.  The dots represent our sample without distance ambiguity, the open triangles sources at near kinematic distance and the filled triangles sources at the far kinematic distance. The pentagons show data from \citet{bontemps 1996} and the squares data from \citet{cabrit 1992}. Open squares and pentagons represent class I sources and filled symbols class 0 sources. \label{relations}}
\end{figure*}

Fig. \ref{relations} shows the relations between the mechanical
force $F_{\rm{CO}}$, the core mass $M_{\rm{core}}$ and the bolometric
luminosity $L$. $F_{\rm{CO}}$ is derived from $F_{\rm{m}}$ (Table
\ref{output}) by applying an inclination correction. Because we do not
know the inclination for each source separately we apply a mean
correction factor of 3, which corresponds to a mean inclination angle
of $57.3^{\circ}$ \citep{bontemps 1996}. The main result of this
comparison is that well established energetic correlations for
low-mass outflows show a continuity up to the high-mass regime. A similar
correlation was already presented by \cite{lada 1985}. 

The low-mass data can be divided into class 0 and class I
sources, and Fig. \ref{relations}(b) suggests tentatively that the
energetic parameters in the high-mass regime correlate better with
class I sources. This might be interpreted as an indication that maybe
no proper class 0 stage exists for massive stars. But clearly this
proposition has to be checked for larger samples to draw firm
conclusions.

Nevertheless, the continuity between the low-mass and the high-mass
regime is strongly suggestive of an outflow formation mechanism which
is present both in the low-mass as well as in the high-mass case.

 \subsection{Mass entrainment and accretion rates}
\label{rates}

\begin{figure}
\includegraphics[angle=-90,width=8cm]{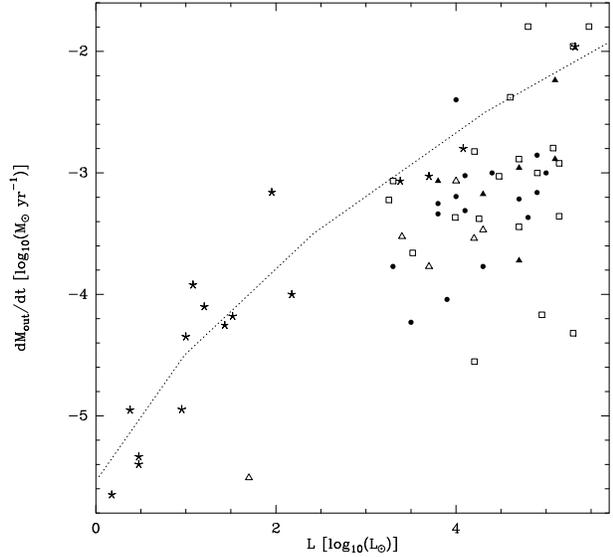}
\caption{Mass entrainment rates versus luminosities (from \citealt{sridha}). The dots (no distance ambiguity), open and filled triangles (near and far kinematic distance, respectively) show the data presented in this paper, the squares represent the data compiled from \cite{churchwell 2000b} and the asterisks the data from \citet{cabrit 1992}. The dotted line shows the second order polynomial fit to the data of \citet{cabrit 1992} as proposed by \citet{shepherd 1996b}. \label{accr_lum}}
\end{figure}

Fig. \ref{accr_lum} presents an update of the mass entrainment rate
versus bolometric luminosity plot presented initially by
\cite{shepherd 1996b}. In addition to our data we also include the 
massive outflow database compiled by \cite{churchwell 2000b}. Although
the spread in the literature data is larger than for our more
homogenous sample the mean of both is similar. This updated plot shows
that the relation fitted by \cite{shepherd 1996b} to the data of
\citet{cabrit 1992} is just an upper envelope in the high-mass
regime. The data indicate that the mass entrainment rate for sources
$>1000$~L$_{\odot}$ does not depend strongly on the luminosity, and
has an average $\dot{M}_{\rm{out}}=\rm{d}M_{\rm{out}}/\rm{d}t$ of
$7.5\times 10^{-4}$~M$_{\odot}$yr$^{-1}$ with a spread between
$10^{-4}$~M$_{\odot}$yr$^{-1}$ and $10^{-3}$~M$_{\odot}$yr$^{-1}$
(these numerical values are calculated just for our sample). 
While part of the non-correlation between $\dot{M}_{\rm{out}}$ and $L$
at high luminosities is certainly due to a variety of observational
uncertainties, we believe that much of the dispersion is real. One
must remember that while outflows in low-mass star formation regions
can usually be identified with a given proto-star, in the regions of
high-mass star formation studied here, we presumably deal with a
cluster of embedded objects. It is possible that in some of our
regions, for example, the most luminous object is on the ZAMS and has
started nuclear burning whereas the most massive outflow emanates from
a less evolved object. Proving this however will require higher
angular resolution and comparison with VLA data.

More interesting than the mass entrainment rates are the accretion
rates $\dot{M}_{\rm{acc}}$ of such massive objects.  Regarding the
flows as momentum driven the momenta of the observed outflow and the
internal jet entraining the outflow should be conserved, if there is
efficient mixing at the jet/molecular gas interface assuming no loss
of momentum to the ISM \citep{richer 2000}:
\begin{eqnarray}
p_{\rm{flow}} = M_{\rm{out}}v_{\rm{out}} = \dot{M}_{\rm{jet}}v_{\rm{jet}}t =
p_{\rm{jet}}. \label{momentum}
\end{eqnarray}
The mean value $v_{\rm{flow}}$ of the observed
$v_{\rm{max}}$ (Table \ref{input}) is $\sim 15$~km/s. Correcting this
by a mean inclination angle of $57^{\circ}$ (multiply by
$1/\cos(57^{\circ})$) results in a molecular outflow velocity of
approximately 28~km/s. This is still a lower limit because we are
noise-limited, and material at higher velocities is likely.

\begin{figure}[ht]
\includegraphics[angle=-90,width=8cm]{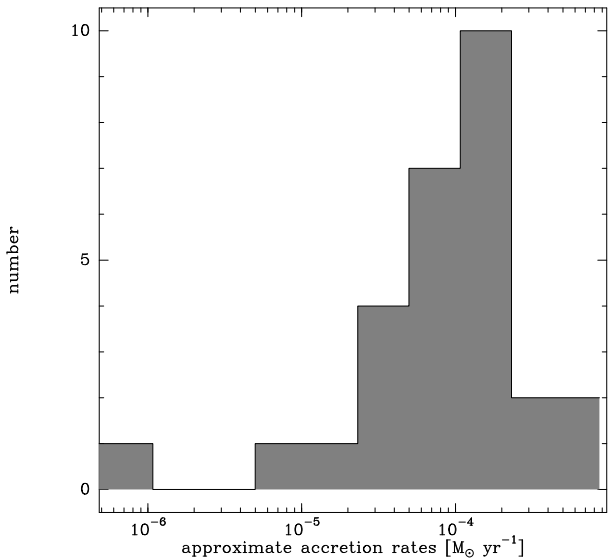}
\caption{Histogram of the estimated accretion rates for our sample. \label{histo_accr}}
\end{figure}

Jet velocities from high-mass proto-stars have only been determined in
a few cases. \cite{marti 1998} observe from proper motions
jet velocities $v_{\rm{jet}}$ of a young massive star around 500~km/s,
and \cite{eisloeffel 2000} report that values between 500~km/s and
1000~km/s are typical. Based on this we assume in the following a mean
ratio between jet velocity and molecular outflow velocity around 20,
which results in jet masses $M_{\rm{jet}}$ and mass loss rate of the
jets $\dot{M}_{\rm{jet}}$ about an order of magnitude below the rates
of the entrained gas. Assuming further a ratio between mass loss rate
of the jet and accretion rate of approximately 0.3
\citep{tomisaka 1998,shu 1999}, we get a mean accretion rate of $\sim
10^{-4}$~M$_{\odot}$yr$^{-1}$ for sources in the $10^4$~L$_{\odot}$
regime ; the distribution is shown in Figure
\ref{histo_accr}. For the most luminous sources ($\sim
10^5$~L$_{\odot}$) listed by \cite{churchwell 2000b} the accretion
rates get as high as $10^{-3}$~M$_{\odot}$yr$^{-1}$. Such values are
extremely high compared with typical accretion rates established for
low-mass star formation ($10^{-6}-10^{-5}$~M$_{\odot}$yr$^{-1}$,
\citealt{shu 1977}). High mass accretion rates moreover ease the 
problem of overcoming radiation pressure and are required in order to
form massive stars within a free-fall time \citep{jijina 1996}. It
will be useful to examine the birth-line for such high accretion rates
in more detail (see, e.g., \citealt{stahler 2000,norberg 2000}).

Other interpretations are still possible, and we cannot rule out that
the chaotic sources without clear outflow structure could be generated
by interacting proto-stars. Collisions of proto-stars are expected to be
very energetic \citep{stahler 2000} causing either real explosions
during stellar merger, or in less dramatic scenarios they could
significantly affect the rotation axis 
of the proto-stellar disks and thus cause strong
precession of the outflows. But bipolar outflows
are very difficult to explain in massive star formation
scenarios based on the coalescence of proto-stars
\citep{bonnell 1998,stahler 2000}. In contrast, high accretion 
rates as found in this study give strong support to theories, which
explain massive star formation mainly by accretion analogous to
low-mass scenarios albeit with accretion rates which are significantly
higher. The fact that highly collimated and massive flows are found
(this work and studies with higher resolution by \citet{beuther 2002b}
and Wyrowski et al. in prep.) supports the proposition that the
physical processes in the low- as well as the high-mass regime are
similar.

\subsection{Outflow mass versus core mass}

\begin{figure*}[ht]
\includegraphics[angle=-90,width=17cm]{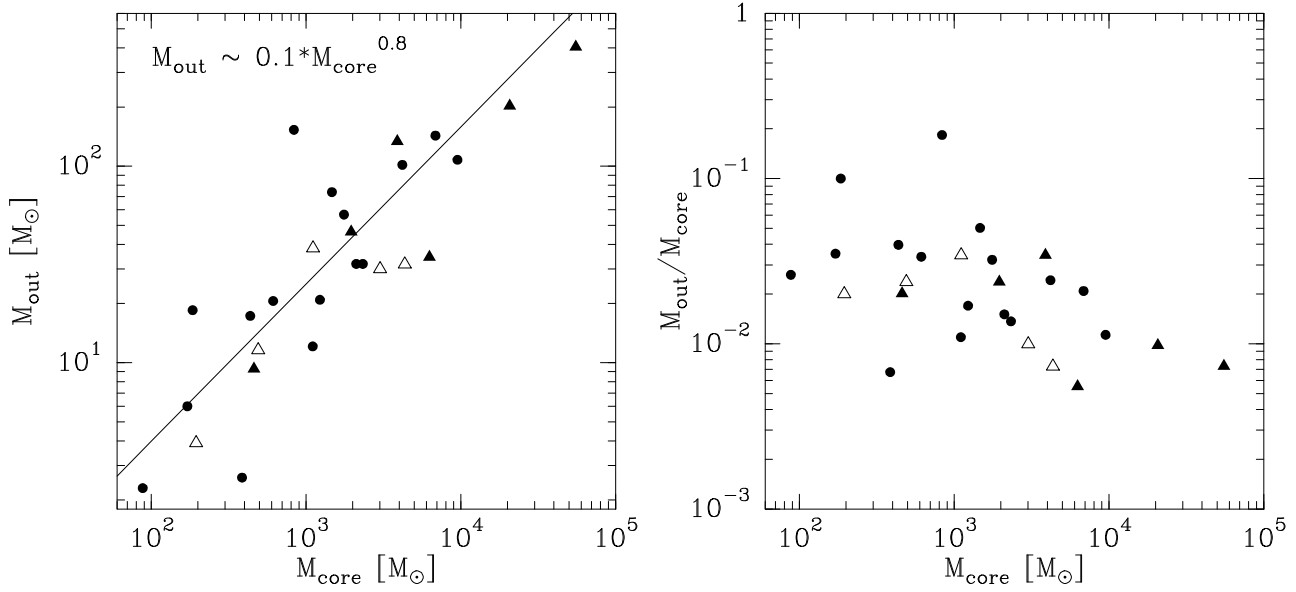}
\caption{The left plot shows the molecular outflow mass versus core mass \citep{beuther 2002a}, whereas the right plot presents the ratio of both quantities versus the core mass. The circles show sources with no distance ambiguities, the filled and open triangles represent sources with far and near distances, respectively. \label{mass_out}}
\end{figure*}

We find a good correlation between the outflow mass $M_{\rm{out}}$ and
the core mass $M_{\rm{core}}$ derived from the 1.2~mm dust emission
\citep{beuther 2002a}, which is presented in Fig. \ref{mass_out}
(left). Our best fit over more than 3 orders of magnitude in core mass
is $M_{\rm{out}}\sim 0.1\cdot M_{\rm{core}}^{0.8}$. The ratio
$M_{\rm{out}}/M_{\rm{core}}$ is rather constant with an average of
0.04 for the sources (Figure
\ref{mass_out} right) and a spread of less than one order of
magnitude.  Thus, approximately $4\%$ of the core gas is entrained in 
the molecular outflow. There is a trend in our data that the
ratio $M_{\rm{out}}/M_{\rm{core}}$ decreases with rising core mass
$M_{\rm{core}}$, but the statistics at the upper mass end are too low
for a definite conclusion on this point.

From the correlation between outflow and core mass, we estimate the
accretion efficiency $f_{\rm{acc}}$. As discussed in \S
\ref{morphologies} the derived parameters include contributions from
several flows, but they are most likely dominated by one massive flow
(see also the discussion in \citealt{tan 2002}),
and we neglect effects due to multiple sources in the following. We
define $f_{\rm{acc}}$
\begin{eqnarray}
f_{\rm{acc}}=\frac{\dot{M}_{\rm{acc}}}{M_{\rm{core}}/t_{\rm{ff}}} \label{accr_eff}
\end{eqnarray}
with the accretion rate $\dot{M}_{\rm{acc}}$ onto the star powering
the massive flow observed by us, the free-fall time $t_{\rm{ff}}$ and
the core mass $M_{\rm{core}}$. Furthermore, the ratio between the mass
loss rate of the jet and the accretion rate is
\begin{eqnarray} 
f_{\rm{r}}=\frac{\dot{M}_{\rm{jet}}}{\dot{M}_{\rm{acc}}} \label{fr}
\end{eqnarray} 
Multiplying eqs. \ref{accr_eff} \& \ref{fr} and using $t\approx t_{\rm{ff}}$ (Fig. \ref{tt}) we get:
\begin{eqnarray}
f_{\rm{r}}f_{\rm{acc}}  &=& \frac{\dot{M}_{\rm{jet}}}{M_{\rm{core}}/t_{\rm{ff}}} = \frac{M_{\rm{jet}}}{M_{\rm{core}}} = \frac{M_{\rm{jet}}}{M_{\rm{out}}} \frac{M_{\rm{out}}}{M_{\rm{core}}} \label{xyz}
\end{eqnarray}
The further assumption of momentum conservation (see
eq. \ref{momentum}) results in:
\begin{eqnarray}
f_{\rm{r}} f_{\rm{acc}} &=& \frac{v_{\rm{out}}}{v_{\rm{jet}}} \frac{M_{\rm{out}}}{M_{\rm{core}}} \label{ff2}
\end{eqnarray}
As outlined in \S\ref{rates} we assume $v_{\rm{out}}$/$v_{\rm{jet}}$
to approximately 1/20. Using additionally the average ratio between
$M_{\rm{out}}$ and $M_{\rm{core}}$ of 0.04 Eq. \ref{ff2} reads:
\begin{eqnarray}
f_{\rm{r}} f_{\rm{acc}} &=& \frac{1}{20} \times 0.04 \sim 2 \times 10^{-3}\label{ff3}
\end{eqnarray}
The value $2\times 10^{-3}$ in Eq. \ref{ff3} has to be taken with
caution because the errors in core and outflow mass are both
roughly a factor 2, and it is not clear if the assumption of momentum
conservation is correct or if some momentum is lost to the ISM exterior
to the core observed at 1.2~mm. But it
is interesting that the product $f_{\rm{r}} f_{\rm{acc}}$ does not
change significantly over many orders of magnitude in core and outflow
mass. This implies that the ratio between the ejected jet mass
$M_{\rm{jet}}$ and the core mass $M_{\rm{core}}$ is roughly constant
for all outflows and cores (see eq. \ref{xyz}).

\subsubsection{Empirical estimates of $f_{\rm{r}}$ and $f_{\rm{acc}}$}

\cite{richer 2000} describe an observational approach to estimate 
$f_{\rm{r}}$. They consider two options:

(a) the luminosity is accretion dominated 
\begin{eqnarray}
f_{r} = \frac{F_{\rm{CO}}}{L} v_{\rm{kep}} \frac{v_{\rm{kep}}}{v_{jet}} \label{f1} 
\end{eqnarray}

(b) the luminosity is mainly due to ZAMS stars
\begin{eqnarray} 
f_{r} = \frac{p_{\rm{CO}}}{M_{\ast}v_{\rm{kep}}} \label{f2}
\end{eqnarray}
with $v_{\rm{kep}}$ the keplerian speed at the proto-stellar
surface (equal to $\sqrt{GM_{\ast}/R_{\ast}}$), which we assume to be
the same as the jet velocity; $F_{\rm{CO}}$ and $p_{\rm{CO}}$ are the
mechanical luminosity and momentum from Table \ref{output} multiplied
by the above discussed statistical inclination correction factor 3,
and $M_{\ast}$ and $R_{\ast}$ are the stellar masses and radii for the
ZAMS \citep{lang}. As luminosities, we use the bolometric
luminosities presented in Table \ref{input}.

The errorbars in both estimates are large (see also \citealt{richer
2000}). Masses and radii of the massive proto-stars are poorly
constrained, and the momenta and mechanical luminosities are uncertain
to about an order of magnitude as well \citep{cabrit 1992}. Thus, the
estimated errors in Eq. \ref{f1} \& \ref{f2} are larger than one
order of magnitude. Additionally, the spread in the derived values of
$f_{r}$ is about an order of magnitude for both approaches. Hence,
the derived parameters should be taken very cautiously, but
nevertheless, Eqs. \ref{f1} \&
\ref{f2} give approximate values for $f_{\rm{r}}$ directly 
from the observations. Additionally, we can estimate from
Eq. \ref{ff3} the accretion efficiency $f_{\rm{acc}}$.

Table \ref{ffresults} gives the median values for $f_{\rm{r}}$ and
$f_{\rm{acc}}$ taken via approach (a) and (b). Both scenarios --~the
accretion dominated and the ZAMS dominated ~-- yield $f_{\rm{r}}$
of order 0.1, which is smaller though not greatly
(given the uncertainties) than theoretical estimates of 0.3 by
\cite{tomisaka 1998} or \citet{shu 1999}.
\cite{richer 2000} find mean values $f_{\rm{r}}$ around 0.3 in the accretion 
dominated case and values about an order of magnitude lower in the ZAMS
case, which is consistent with our results.

\begin{table}[ht]
\caption{Median $f_{\rm{r}}$ and $f_{\rm{acc}}$: (a) luminosity is accretion dominated; (b) luminosity is ZAMS dominated \label{ffresults}}
\begin{tabular}{lcc}
    & $f_{\rm{r}}$ & $f_{\rm{acc}}$\\
\hline
(a) & 0.17 & 0.01 \\ 
(b) & 0.07 & 0.03
\end{tabular}
\end{table}

Rephrasing Eq. \ref{accr_eff} results in the accretion rate being
a rather linear function of the core mass with the accretion
efficiency and the free-fall timescale ($\sim 10^5$~yr) being
approximately constant over the whole mass range. In the
accretion dominated case ($f_{\rm{acc}}\sim 0.01$, see also
\cite{sridha}) we find:
\begin{eqnarray}
\dot{M}_{\rm{acc}} = \frac{f_{\rm{acc}}}{t_{\rm{ff}}} M_{\rm{core}} = 1 \times 10^{-7} \times M_{\rm{core}}\ [\rm{M}_{\odot} \rm{yr}^{-1}]. \label{accr}
\end{eqnarray}
Using additionally $M_{\ast}=\dot{M}_{\rm{acc}}\times t$ for the mass of the 
forming star, we get:
\begin{eqnarray}
M_{\ast} = f_{\rm{acc}} M_{\rm{core}} \label{star}
\end{eqnarray}
As already stressed, the possible numerical uncertainties are large
(about an order of magnitude), but equations
\ref{accr} \& \ref{star} result, e.g., for a 5000~M$_{\odot}$ core, 
in an accretion rate of $\sim 5\times 10^{-4}$~M$_{\odot}$/yr and a
stellar mass of 50~M$_{\odot}$ for the most massive object of the
evolving cluster. These are plausible values, and the accretion rate
corresponds well to the range of accretion rates derived in \S
\ref{rates}. Such accretion rates are sufficiently high to overcome the
radiation pressure and form massive stars (see, e.g., \citealt{jijina
1996}). Additionally, relation \ref{star} agrees reasonably well with
estimates of the most massive star of a cluster assuming a star
formation efficiency of $30\%$ \citep{lada 1993} and a Miller-Scalo
IMF \citep{miller 1979}. We conclude that one can explain the
formation of stars of all masses by similar accretion based scenarios
with accretion rates roughly proportional to final masses
\citep{norberg 2000,bonnell 2001, tan 2002}.

%%%%%%%%%%%%%%%%%%%%%%%%%%%%%%%%%%%%%%%%%%%%%%%%%%%%%%%%%%%%%%%%%%%%%%%%%
% SUMMARY
%%%%%%%%%%%%%%%%%%%%%%%%%%%%%%%%%%%%%%%%%%%%%%%%%%%%%%%%%%%%%%%%%%%%%%%%%

\section{Summary}

CO mapping of 26 sources showing line wing emission in previous single
pointing observations reveal a high degree of bipolar outflow
morphology with 21 sources being resolved into bipolar
structures. This supports the idea that bipolar outflows are not only
associated with low-mass star formation, but are ubiquitous phenomena
for all masses. The size of the flows is on the pc-scale, and the data
reveal collimation factors on the average $\geq 2$, which is similar
to results for low-mass flows but differs from previous high-mass
estimates in the literature \citep{richer 2000}. We conclude that
``Orion--type'' flows of low collimation factor are either rare in
general or confined to very massive proto-stars (see however the
discussion by \citealt{reipurth 2001}).

The dynamical timescales of the outflows
correspond well to the free-fall timescales of the associated cores.
This suggests that for high-mass flows, the CO timescale may
often be a good estimate of the flow age. 

Derived flow parameters show that we are really dealing with 
massive and energetic sources, and comparing those quantities with
equivalent studies in the low-mass regime shows that previously
derived correlations for low-mass proto-stars have counterparts in the
high-mass regime. We interpret that as support for star formation
scenarios, which predict similar physical accretion processes for all
masses with significantly increasing accretion rates
$\dot{M}_{\rm{acc}}$ in high-mass cores.

We show that mass entrainment rates increase with core luminosity, and
most of the massive sources show mass entrainment rates between $\sim
10^{-4}$~M$_{\odot}$/yr and $\sim 10^{-3}$~M$_{\odot}$/yr with a few
cases of higher values. In momentum driven flows, these mass
entrainment rates should correspond to mean accretion rates of order
$10^{-4}$~M$_{\odot}$yr$^{-1}$ for sources of bolometric luminosity
$\sim 10^4$~L$_{\odot}$. This is in good agreement with recent massive
star formation scenarios, which predict higher accretion rates for the
most massive stars
\citep{norberg 2000,bonnell 2001,tan 2002}.

We find a tight correlation between the outflow mass and the core
mass, which holds over more than three orders of magnitude in core
mass. The ratio of both quantities is rather constant around
0.04. This correlation indicates that the product of the accretion
efficiency and the ratio between the mass entrainment rate and the
accretion rate, which equals the ratio between jet and core mass
$f_{\rm{acc}} f_{\rm{r}}= M_{\rm{jet}}/M_{\rm{core}}$, is roughly
constant during star formation of all masses. This makes the accretion
rate (and thus the mass of the most massive star of the cluster) to
first order a linear function of the core mass.

Estimates of $f_{\rm{r}}$ and $f_{\rm{acc}}$ are around 0.2 and 0.01,
respectively. In spite of large uncertainties, the results are
consistent with current jet entrainment flow formation scenarios
\citep{richer 2000}.

Those results support star formation theories in the high-mass regime
which are based on similar principles as those for low-mass star
formation (e.g., \citealt{norberg 2000,jijina 1996,wolfire 1987,tan
2002}). The intrinsic parameters are more energetic, but the physical
processes are likely to be similar.

%%%%%%%%%%%%%%%%%%%%%%%%%%%%%%%%%%%%%%%%%%%%%%%%%%%%%%%%%%%%%%%%%%%%%%%%%%
% acknoledgements
%%%%%%%%%%%%%%%%%%%%%%%%%%%%%%%%%%%%%%%%%%%%%%%%%%%%%%%%%%%%%%%%%%%%%%%%%%

\begin{acknowledgements} 
We like to thank the unknown referee for helpful and improving
comments on the initial manuscript. H. Beuther gets support by the
{\it Deutsche Forschungsgemeinschaft, DFG} project number SPP
471. F. Wyrowski is supported by the National Science Foundation under
Grant No. AST-9981289.
\end{acknowledgements}

%%%%%%%%%%%%%%%%%%%%%%%%%%%%%%%%%%%%%%%%%%%%%%%%%%%%%%%%%%%%%%%%%%%%%%%%%%
% bibliography
%%%%%%%%%%%%%%%%%%%%%%%%%%%%%%%%%%%%%%%%%%%%%%%%%%%%%%%%%%%%%%%%%%%%%%%%%%

\end{document}